\documentclass[modern]{aastex63}

\usepackage{xspace}
\usepackage{cancel}
\usepackage{acro}
\acsetup{
  single    = false,
  list/sort  = true,
  cite/group = true,
  cite/group/cmd = \citealp,
  cite/group/pre = {;\xspace},
}

\newcommand{\mearth}{\mbox{$\mathrm{M}_{\oplus}$}\xspace}

\DeclareAcronym{tess}{
  short = TESS ,
  long = Transiting Exoplanet Survey Satellite,
  cite = {ricker15}
}

\DeclareAcronym{ecmi}{
  short = ECMI,
  long = electron cyclotron maser instability
  }
  
\DeclareAcronym{jwst}{
  short = JWST ,
  long = James Webb Space Telescope,
  cite= {jwst}
  }
  
\DeclareAcronym{askap}{
  short = ASKAP,
  long = the Australian Square Kilometre Array Pathfinder
  }
  
\DeclareAcronym{rms}{
  short = RMS,
  long = root-mean-square
  }
  
\DeclareAcronym{zdi}{
  short = ZDI,
  long = Zeeman Doppler Imaging
  }
  
\DeclareAcronym{lotss}{
  short = LoTSS,
  long = The LOFAR Two-metre Sky Survey,
  cite = {shimwell17}
}

\DeclareAcronym{hpf}{
  short = HPF,
  long = the Habitable-Zone Planet Finder,
  cite = {hpf}
}

\DeclareAcronym{spi}{
  short = SPI,
  long = star-planet magnetic interactions,
  cite = {zarka2001}
}

\DeclareAcronym{racs}{
  short = RACS,
  long = Rapid ASKAP Continuum Survey,
}

\DeclareAcronym{rv}{
  short = RV,
  long = radial velocity
}
\DeclareAcronym{cme}{
  short = CME,
  long = coronal mass ejections
}

\definecolor{linkcolor}{rgb}{0.1216,0.4667,0.7059}

\usepackage{xcolor, fontawesome}
\definecolor{twitterblue}{RGB}{64,153,255}
\newcommand\twitter[1]{\href{https://twitter.com/#1 }{\textcolor{twitterblue}{\faTwitter}\,\tt \textcolor{twitterblue}{@#1}}}
\usepackage{amsmath}

\graphicspath{{../notebooks/}}

\shorttitle{The TESS View of LOFAR Radio-Emitting Stars}
\shortauthors{B. J. S. Pope et al.}

\begin{document}

\title{The TESS View of LOFAR Radio-Emitting Stars}

\correspondingauthor{Benjamin J. S. Pope \twitter{fringetracker}}
\email{b.pope@uq.edu.au}

\author[0000-0003-2595-9114]{Benjamin J. S. Pope}
\affiliation{School of Mathematics and Physics, The University of Queensland, St Lucia, QLD 4072, Australia}
\affiliation{Centre for Astrophysics, University of Southern Queensland, West Street, Toowoomba, QLD 4350, Australia}

\author[0000-0002-7167-1819]{Joseph R. Callingham}
\affiliation{Leiden Observatory, Leiden University, PO Box 9513, 2300 RA, Leiden, The Netherlands}
\affiliation{ASTRON, Netherlands Institute for Radio Astronomy, Oude Hoogeveensedijk 4, Dwingeloo, 7991 PD, The Netherlands}

\author[0000-0002-9464-8101]{Adina D. Feinstein}
\affiliation{Department of Astronomy and Astrophysics, University of Chicago, 5640 S. Ellis Ave, Chicago, IL 60637, USA}
\affiliation{NSF Graduate Research Fellow}

\author[0000-0002-3164-9086]{Maximilian N. G\"{u}nther}
\affiliation{Department of Physics, and Kavli Institute for Astrophysics and Space Research, Massachusetts Institute of Technology, Cambridge, MA
02139, USA}
\affiliation{Juan Carlos Torres Fellow}

\author[0000-0002-0872-181X]{Harish K. Vedantham}
\affiliation{ASTRON, Netherlands Institute for Radio Astronomy, Oude Hoogeveensedijk 4, Dwingeloo, 7991 PD, The Netherlands}
\affiliation{Kapteyn Astronomical Institute, University of Groningen, PO Box 72, 97200 AB, Groningen, The Netherlands}

\author[0000-0003-4142-9842]{Megan Ansdell}
\affiliation{NASA Headquarters, 300 E Street SW, Washington, DC 20546, USA}

\author[0000-0001-5648-9069]{Timothy W. Shimwell}
\affiliation{ASTRON, Netherlands Institute for Radio Astronomy, Oude Hoogeveensedijk 4, Dwingeloo, 7991 PD, The Netherlands}
\affiliation{Leiden Observatory, Leiden University, PO Box 9513, 2300 RA, Leiden, The Netherlands}

\begin{abstract}
The recent detection of the M~dwarf GJ\,1151 at 144\,MHz low radio frequencies using LOFAR has been interpreted as evidence of an exoplanet magnetically interacting with its host star. This would be the first exoplanet detected around a main sequence star by a radio telescope. Radial velocity confirmation of such a planet has proven inconclusive, and it remains possible that the radio emission could be generated by a stellar coronal process. Using data from TESS, we shed light on this question by probing the stellar activity and flares of GJ\,1151 as well as fourteen other M~dwarfs detected by LOFAR. GJ\,1151 and three other star-planet interaction candidates are found to be inactive, with no rotational modulation and few, if any, flares. The remainder of the LOFAR detected M~dwarfs flare frequently. We consider it unlikely that stellar activity is responsible for the bright, circularly-polarized emission from GJ\,1151 and its three analogs, and support the star-planet magnetic interaction interpretation. \href{https://github.com/benjaminpope/mtv}{\color{linkcolor}\faGithub} 
\newpage
\end{abstract}


\section{Introduction} 
\label{sec:intro}

Since the dawn of radio astronomy, it has been known that the Sun and planets of our Solar System are bright sources of radio emission \citep{pawsey46,kaiser84}. 
The Sun emits radio waves from its active regions, flares, and corona \citep{melrose80,dulk85}, while the Earth and planets produce low-frequency auroral radio emission \citep{zarka2001}. In addition to this, Jupiter electrodynamically interacts with its innermost moons, especially Io, giving rise to  strong coherent radio emission \citep{bigg65}.

Nevertheless, outside of the Solar System, the vast majority of detected low-frequency radio sources are active or star-forming galaxies, compact objects, or interacting binaries. Main-sequence stars detected in the radio are typically nearby ($\lesssim 100\,{\rm pc}$) and chromospherically active \citep{guedel93}. While stellar flares can be bright in the radio spectrum \citep{crosley18,zic20}, only a handful of the nearest main-sequence stars have been detected in quiescent emission \citep[e.g.][]{trigilio18,perez21}. Detecting coherent radio waves from quiescent stellar systems would provide valuable information not only about the magnetospheres of the vast majority of stars, but also about possible \ac{spi}. \citet{perez21} interpret variability in the radio emission of Proxima~Cen as SPI, though this is inconsistent with other models \citep{kavanagh21}.
While known exoplanet-host stars have been targeted by sensitive radio searches, no exoplanets have been conclusively identified in low-frequency radio emission \citep{lynch18,turner21}.  

Radio emission is a strong indication of the space weather environment, which is key to understanding planetary habitability. M~dwarf stars are thought to host the great majority of temperate terrestrial planets \citep{hsu20}, defined such that equilibrium temperatures could sustain liquid water at the planet's surface. Around small, dim stars, planetary transits are deep ($\sim 1-10\%$) and the liquid-water habitable zone contains short orbital periods on the order of days. 
The potential for high-quality atmospheric characterization of habitable planets using transmission spectroscopy \citep{morley17} means that M~dwarfs are a cornerstone of the transiting exoplanet science planned with \ac{tess} and \ac{jwst}. 


M~dwarfs are highly active, especially when they are young and rapidly rotating \citep{newton17}, and close-in `habitable' planets may be rendered uninhabitable by a harsh space weather environment and frequent flares \citep{tarter07,shields16}. A risk is that the liquid-water habitable zone may lie inside of the Alfv\'{e}n surface, defined such that the Alfv\'{e}n speed exceeds the plasma flow speed. Within this region the magnetic field lines threading a planet's plasma environment are directly connected to the field of the star, rather than (as for Earth) forming separate magnetospheres delineated by a shock discontinuity. The stellar wind can then impinge directly on the planet's atmosphere, potentially stripping it of volatiles and irradiating the surface in a way that would be hostile to life \citep{garraffo17}. 
For these reasons, detecting radio emission from \ac{spi} is rare observational probe of the stellar influence on M~dwarf planetary habitability \citep{kavanagh21}.

\subsection{M~Dwarf Radio Emission}
\label{sec:mdwarf_radio}

Radio emission from stellar systems can be classified into coherent or incoherent processes, distinguished by radiation characteristics such as the brightness temperature, degree of circular polarisation, and duration \citep{melrose80,dulk85}. 

Coherent radio emission processes are further divided two categories: plasma and \ac{ecmi} emission. Plasma emission can be driven by stellar activity, produced via impulsively heated plasma being injected into a colder plasma \citep{stepanov2001,osten06} as in stellar flares and \ac{cme} \citep{dulk85,matthews19}. We refer to nonthermal radio emission from plasma distributions heated by stellar flares and magnetic reconnection as `activity-driven'.

\ac{ecmi} emission can also be generated via auroral processes similar to those seen on Jupiter \citep{zarka1998}. Currents that accelerate electrons into the neutral Jovian atmosphere can be established by the breakdown of co-rotation when the magnetospheric plasma lags behind the magnetic field of Jupiter \citep{hill1976}, or by Jupiter's magnetic field sweeping over its conducting satellite Io \citep{turnpenney17}. The loss of high-energy electrons in the atmosphere of Jupiter establishes the population inversion necessary for ECMI emission \citep{cowley2001}. We define stellar auroral emission in this context as any stellar \ac{ecmi} radio emission that is not generated directly via stellar activity, with Jovian analog processes in mind.

\subsection{LOFAR Radio Detections}

Instead of using a targeted radio study as in previous detections, a larger number of radio-bright M~dwarfs have now been detected using wide-field interferometric surveys. For example, using \ac{askap} at 888\,MHz in the Rapid \ac{askap} Continuum Survey of the whole sky $\delta < +41^\circ$, \citet{pritchard21} have identified emission from 33 stars, including chemically-peculiar stars, interacting binaries, and 18 K and M~dwarfs. 

At lower frequencies still, 
LOFAR \citep[LOw-Frequency ARray;][]{lofar}, with its deep sensitivity and fast survey speed, is an ideal instrument for searching for this. Low frequencies are the ideal spectral window to search for coherent radio emission associated with star-planet magnetic interactions: the expected \ac{ecmi} emission is cut off above a frequency proportional to the magnetic field strength of the emitter \citep{treumann06}, and can be distinguished from fundamental plasma emission, for which the brightness temperature $T_{B}$ $\propto$ frequency $\nu^{-2}$. \ac{lotss} is an ongoing survey of the entire Northern sky, which currently covers $\approx$20\% of the northern hemisphere using a low-frequency radio band centered at 144\,MHz, reaching sensitivities with $\lesssim$100$\,\mu$Jy \ac{rms} noise an order of magnitude deeper than comparable previous low-frequency surveys. 

Cross-matching LOFAR-detected sources against the Gaia~DR2 optical catalog \citep{gaiadr2} has led to the detection of GJ\,1151 \citep{vedantham20}, a quiescent, slowly-rotating M7 dwarf only 8\,pc away. While it is faint in X-rays \citep{foster20}, it emits bright, highly-circularly-polarized low-frequency radio emission, which \citet{vedantham20} interpret as a sign of \ac{spi} with a terrestrial-mass planet in a few-day orbit. Nevertheless, \ac{rv} observations of GJ\,1151 place limits of 1.2\mearth on any planet with an orbital period shorter than five days \citep{pope20rv,perger21,mahadevan21}. Without confirmation of a planetary orbit or radio modulation at its period, it is not certain whether the radio emission observed from GJ\,1151 is powered by \ac{spi}.

Expanding the LoTSS-Gaia M~dwarf sample, \citet{callingham_pop} have reported the detection of~18 M~dwarfs in addition to GJ\,1151, across a range of spectral types, stellar activity levels, and including both binaries and single stars. They conclude that there may be two sets of emission mechanisms responsible for the low-frequency radio detections: 1) an activity-driven plasma mechanism responsible for polarized bursts, in which solar-like coronal processes are at work in active stars; and 2) an auroral emission mechanism operating in the most inactive stars, with \ac{ecmi} produced by the breakdown of co-rotation between the star and plasma in its magnetosphere for the fastest rotating stars, or by the magnetic interaction with an exoplanet.

\section{The TESS View}

Optical light curves can help distinguish between activity-driven and auroral radio emission mechanisms for these stars. \citet{pineda17} support the idea that radio emission from ultracool dwarfs is auroral partly on the basis that they have much lower flare rates than main-sequence M~dwarfs, and that flares are responsible for the coronal heating which gives rise to the coronal radio emission. On the same basis, the non-detection of optical flares in \ac{spi} candidate systems would add to the evidence that their radio emission source is not coronal. 

In this paper we examine optical light curves of the \citet{callingham_pop} sample using \ac{tess}. Since its launch in 2018, \ac{tess} has been obtaining time-series photometry of nearly the entire sky, in sequential Sectors each of duration 27\,days. 
Fifteen of the LOFAR-detected stars were observed (not simultaneously) with \ac{tess} at 2-minute cadence, high-precision photometry \citep[for $T_\text{mag}\lesssim 10$, typically better than 1\,mmag in 1\,hr;][]{handberg21}, which can reveal flares and starspot modulation from stellar rotation. By applying the \texttt{stella} machine-learning code to detect stellar flares \citep{feinstein20,feinstein20joss}, we obtain uniform measures of flare rate and intensity across the sample, previously-published only for CR\,Dra \citep{callingham_crdra}. 

Two sources, GJ\,450 and 2MASS~J09481615+5114518, were observed simultaneously by LOFAR and TESS, for eight-hour windows beginning UTC 20:11:00 on 2020-03-16 and UTC 20:45:40 on 2020-01-31 respectively. The stars were not detected by LOFAR in this time period, and no flares were detected by TESS during the simultaneous observations, or in the hours immediately before or after.

While sensitive non-detection of optical flares does not necessarily indicate auroral processes, and the observations are not simultaneous, a lack of \ac{tess} flares reveals a quiescent chromosphere where the radio emission is more likely powered by magnetospheric acceleration mechanisms instead of chromospheric ones. We show that the active systems identified by \citet{callingham_pop} display flares in the TESS band as well, while four systems that were identified as quiescent and potentially auroral flare either infrequently or even not at all during the TESS observations.

\begin{deluxetable}{cccccc}
\tablecaption{Properties of LOFAR radio M dwarfs observed by TESS, ordered by increasing flare rate. Asterisks denote stars observed simultaneously by LOFAR and TESS. Rotation periods denoted by daggers are from these TESS observations, and others from literature in \citet{callingham_pop}. Flare rates are empirical rates with Poisson uncertainties. \label{table:papertable}}
\tablehead{\colhead{Name} & \colhead{TIC} & \colhead{X-ray Lum.} & \colhead{LOFAR Lum.} & \colhead{Flare Rate} & \colhead{Rotation Period}\\ \colhead{ } & \colhead{ } & \colhead{($10^{28}$ erg/s)} & \colhead{($10^{14}$ erg/s/Hz)} & \colhead{(d$^{-1}$)} & \colhead{(d$^{-1}$)}}
\startdata
GJ 1151 & 11893637 & $0.02$ & $0.63 \pm 0.15$ & $<0.059$ & 125. \\
LP 169-22 & 148673115 & $<0.03$ & $1.03 \pm 0.48$ & $<0.024$ & -- \\
G 240-45 & 307957392 & $0.02$ & $12.3 \pm 1.57$ & $0.0069~(<0.015)$ & -- \\
GJ 625 & 207492082 & $0.04$ & $0.8 \pm 0.09$ & $0.015~(<0.036)$ & 79.8 \\
2M J0948+5114* & 453430899 & $0.28$ & $28.71 \pm 2.27$ & $0.063~(<0.14)$ & -- \\
GJ 450* & 144400022 & $0.66$ & $0.54 \pm 0.2$ & $0.20~(<0.29)$ & 23.0 \\
WX Uma & 252803603 & $0.36$ & $0.45 \pm 0.09$ & $0.23 \pm 0.1$ & 0.780 \\
DO Cep & 414587194 & $0.23$ & $0.92 \pm 0.1$ & $0.27 \pm 0.07$ & 0.410 \\
LP 259-39 & 166597074 & $<18.7$ & $10.11 \pm 2.75$ & $0.29 \pm 0.1$ & 1.7$^\dagger$ \\
LP 212-62 & 392365135 & $0.38$ & $28.34 \pm 1.55$ & $0.35 \pm 0.1$ & 60.8 \\
2M J1433+3417 & 409372963 & $0.83$ & $30.82 \pm 4.88$ & $0.38 \pm 0.1$ & 0.14$^\dagger$ \\
GJ 3861 & 298164374 & $3.36$ & $3.64 \pm 0.57$ & $0.42 \pm 0.09$ & -- \\
DG CVn & 368129164 & $10.72$ & $2.5 \pm 0.8$ & $0.75 \pm 0.2$ & 0.110 \\
CW UMa & 85334035 & $5.37$ & $4.23 \pm 0.44$ & $1.1 \pm 0.2$ & 7.77 \\
CR Dra & 207436278 & $36.65$ & $43.38 \pm 2.46$ & $1.7 \pm 0.1$ & 1.98
\enddata
\end{deluxetable}

\subsection{TESS Data Reduction}
\label{sec:tessdata}


We examine all M~dwarfs in the \citet{callingham_pop} sample using TESS short-cadence data, to look for flares, rotational modulation, and other signatures of stellar activity. Because these are high proper motion sources, all sources are identified by position and proper motion in the TESS Input Catalog \citep[TIC;][]{tic} by manual inspection of the catalog on the Mikulski Archive for Space Telescopes (MAST), with TIC identifications listed in Table~\ref{table:papertable}. 

A similar pipeline to that used in \citet{callingham_crdra} was applied to all stars in the sample. The \texttt{lightkurve} package \citep{lightkurve} is used to query MAST for each TIC in our target list and download 120\,s-cadence light curves. The Pre-search Data Conditioning Simple Aperture Photometry (PDCSAP) flux for all available Sectors is then cleaned of \texttt{nan} values and quality-flagged epochs, and normalized.

If some of our stars of interest have very low activity rates, a confident measure of a very low rate would suggest that coronal processes are unlikely to be producing the low-frequency radio emission. 

We apply a flare-finding algorithm using convolutional neural networks (CNNs) \citep{feinstein20}. This package, \texttt{stella}, uses the average output from an ensemble of CNNs trained on an existing catalog of flares from TESS  \citep{gunther20}. 

We use \texttt{stella} v.0.1.0 \citep{feinstein20joss} to detect flares in the sector-by-sector PDCSAP light curves, providing per-cadence uncalibrated probabilities $\in (0,1)$ of each time sample belonging to a flare. The time-series is then grouped by contiguous chunks into a table of individual flares, within which peaks are identified. 

\begin{figure}
\plotone{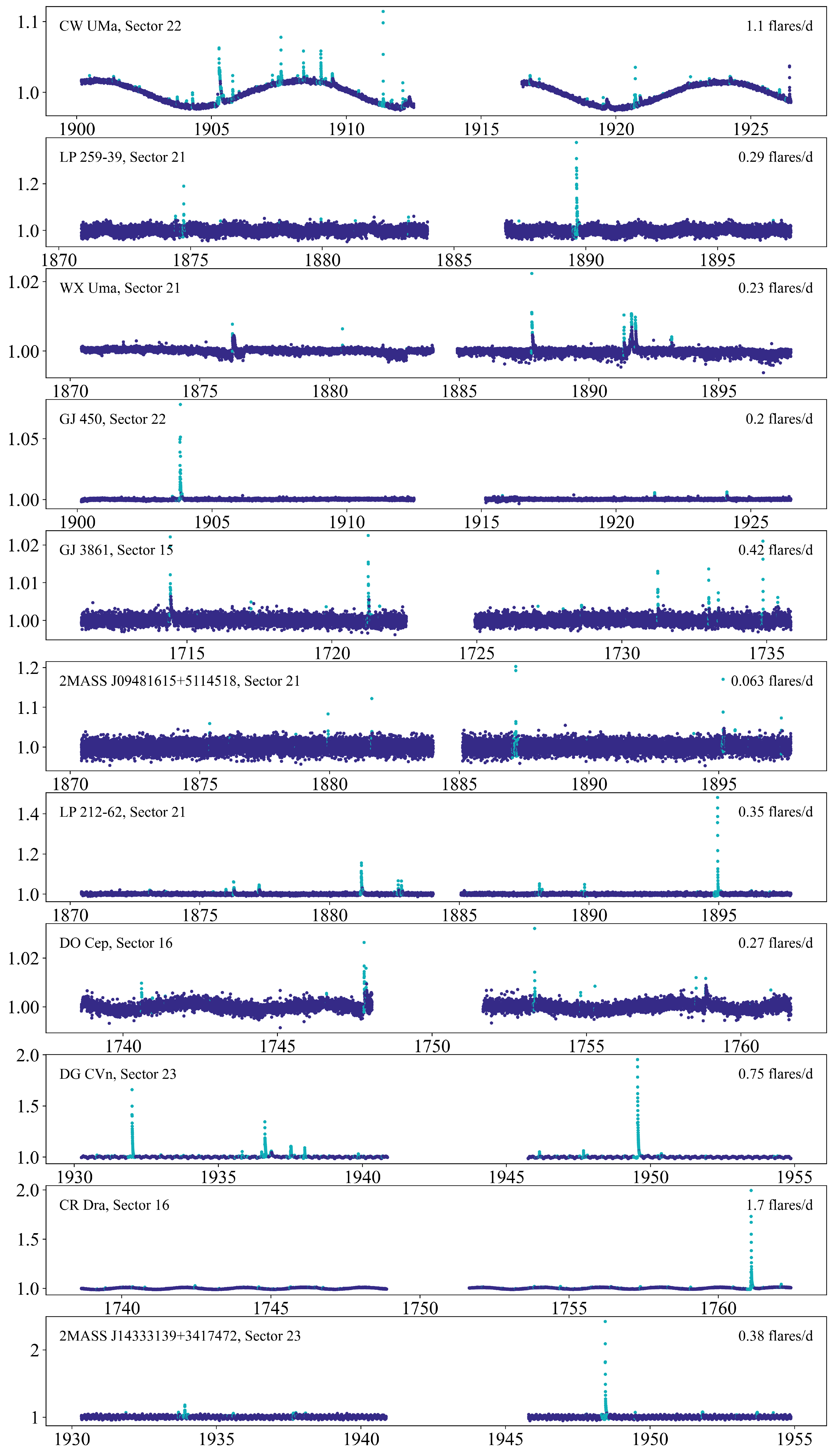}
\epsscale{0.6}
\caption{Single-Sector TESS light curves of the activity-driven candidate sources from the LOFAR radio-emitting sample, with epochs coloured by \texttt{stella} flare probabilities over 0.6 (light blue) and under 0.6 (dark blue). All stars in this subsample are observed to flare significantly. This figure was produced in a Jupyter Notebook, available online. \href{https://github.com/benjaminpope/mtv/blob/master/notebooks/paper_plots.ipynb}{\color{linkcolor}\faGithub} \label{active_stars}}
\end{figure}

\begin{figure}
\plotone{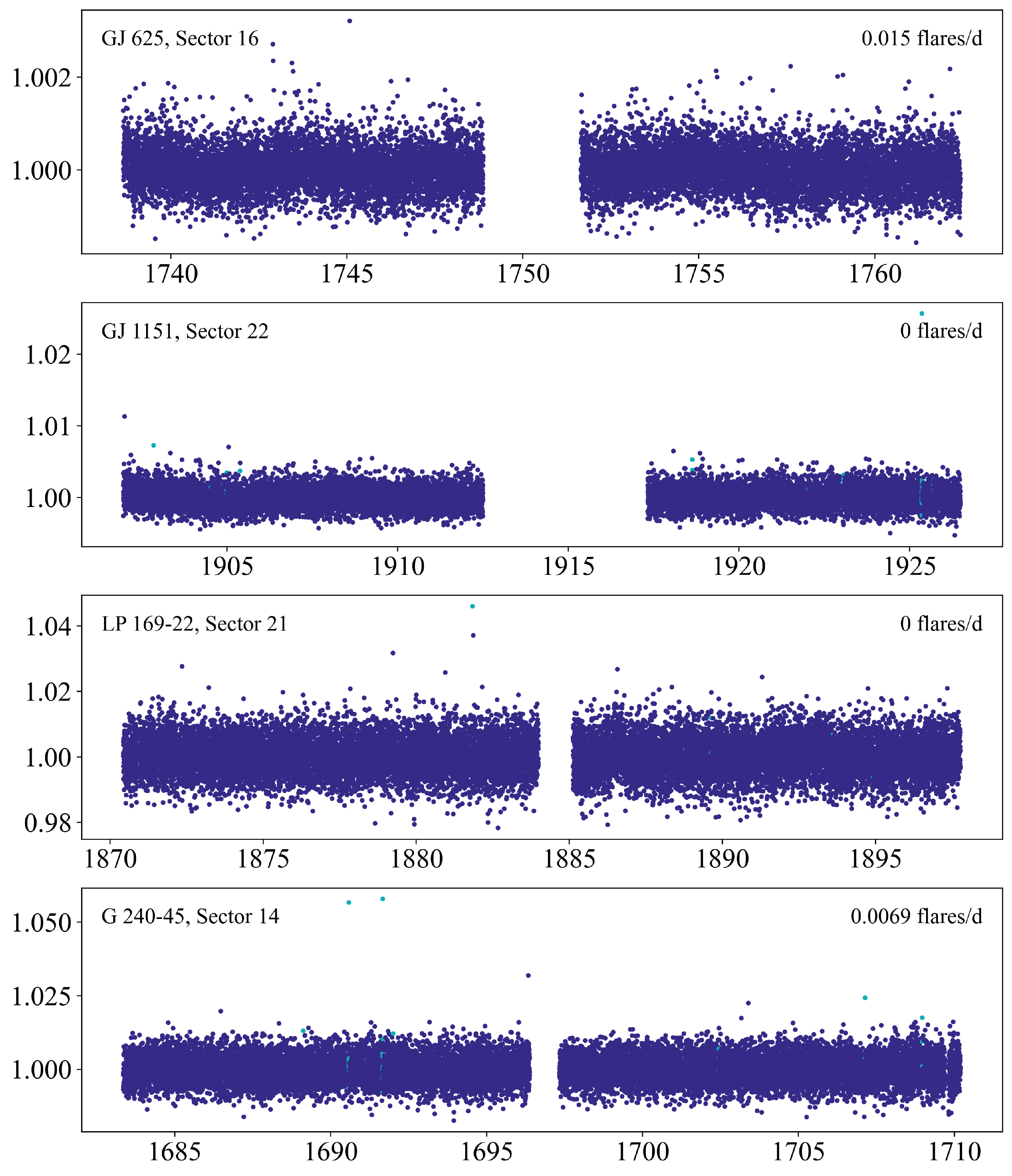}
\caption{Single-Sector TESS light curves of the quiescent auroral emission candidate sources from the LOFAR radio-emitting sample, with epochs coloured by \texttt{stella} flare probabilities over 0.6 (light blue) and under 0.6 (dark blue). The light curves show no periodic modulation in TESS, consistent with low activity levels and long rotation periods ($\gtrapprox 20\,$d). This figure was produced in a Jupyter Notebook, available online.  \href{https://github.com/benjaminpope/mtv/blob/master/notebooks/paper_plots.ipynb}{\color{linkcolor}\faGithub}  \label{quiescent_stars}}
\end{figure}

False positives are identified by simple filters: any flare is removed if its fractional amplitude is lower than $3\times$\,the \ac{rms} of the light curve smoothed on 400~min timescales, or if its fitted duration (rise + fall) is shorter than 4\,minutes, i.e. two TESS cadences. 
The empirical flare rate for each star is then estimated by the probability-weighted sum of the number of flares divided by the observing time. 
Uncertainties are calculated conservatively as two-sided $1\,\sigma$ Poisson confidence intervals \citep{gehrels86}. 
To turn this into a flare rate, we divide this by the total observed time. The results are displayed in Table~\ref{table:papertable}, with light curves coloured by flare classification displayed in Figures~\ref{active_stars} and~\ref{quiescent_stars}. Two of our stars (GJ\,1151 and LP\,169-22) show no flares in TESS, while four more show fewer than 5; for these, we report one-sided $1\,\sigma$ upper limits. 

One highly active source -- WX\,UMa -- is the secondary component of the M1+M6 binary GJ\,412\,AB. With five magnitudes difference in flux between the components, the quiescent primary dominates the light curve. Furthermore, only the primary is allocated 2-min cadence pixels, and as TESS marginally resolves the wide binary, the default SPOC aperture excludes the contribution from the secondary, and no flares are visible. By using a smaller aperture centered on WX UMa, we extract a light curve with a reduced contamination from the brighter primary. The different realization of the systematics here means that \texttt{stella} identifies only one flare, while others can be seen by human inspection but are returned as false negatives in \texttt{stella}. Accordingly we visually examined the light curve in day-long chunks, and manually identified nine flares, and use this number in determining flare rate.

AD Leo lies close to the ecliptic, and is therefore outside the viewing zones in TESS Cycles 1-3. It will, however, be observed in the extended ecliptic mission in Cycle 4, in Sectors 45, 46, and~48.

\section{Discussion}
\label{sec:discussion}

Of the stars considered here, \citet{callingham_pop} identified five targets as being quiescent, whose emission is best understood as auroral: GJ\,450, GJ\,625, GJ\,1151, G\,240-45, and LP\,169-22. These were classified as quiescent based on a conjunction of low H$\alpha$ luminosity, low X-ray luminosity, and long rotation periods (where known). The remainder of the \citet{callingham_pop} sample are all coherent emitters but more active, and emission may be auroral or activity-driven.

The TESS flare rates and rotational variability add a new dimension to this analysis. In Section~\ref{sec:gudel_benz}, we situate these stars in a three-dimensional Güdel-Benz relation between radio luminosity, X-ray luminosity, and TESS flare rate. We find that while GJ\,450 is actively flaring, the four remaining quiescent candidates are very inactive in TESS, and conclude that these are strong candidates for auroral emission and \ac{spi}. 

\subsection{Güdel-Benz Relation}
\label{sec:gudel_benz}
The Güdel-Benz relation is a scaling relation between the X-ray and radio luminosity of many active stars \citep{guedel93,benz94}. This is generally understood to originate from non-thermal radio emission coming from electrons accelerated by magnetic reconnection events associated with stellar flares and prominences; this also heats surrounding plasma in the chromosphere, giving rise to thermal X-ray emission. If radio emission is being generated by this process, we expect not only a high X-ray luminosity, but also a correspondingly high rate of optical flares as the source of the heat driving the X-rays \citep{matthews19}. 

One of the arguments made by \citet{callingham_pop} for why the radio emission from GJ\,1151 and their other quiescent detections must be non-coronal, potentially due to \ac{spi}, is that it disobeys the Güdel-Benz relation: e.g. GJ\,1151 is radio-loud but X-ray faint. Such a violation of the G\"{u}del-Benz relationship is consistent with the radio emission being coherent. 
In the case of auroral emission from a Jupiter-Io process or the breakdown of corotation, we would expect to see radio-bright sources that are under-luminous in X-rays relative to the Güdel-Benz relationship; we would also expect them to have significantly lower flare rates. 

\begin{figure}
\plotone{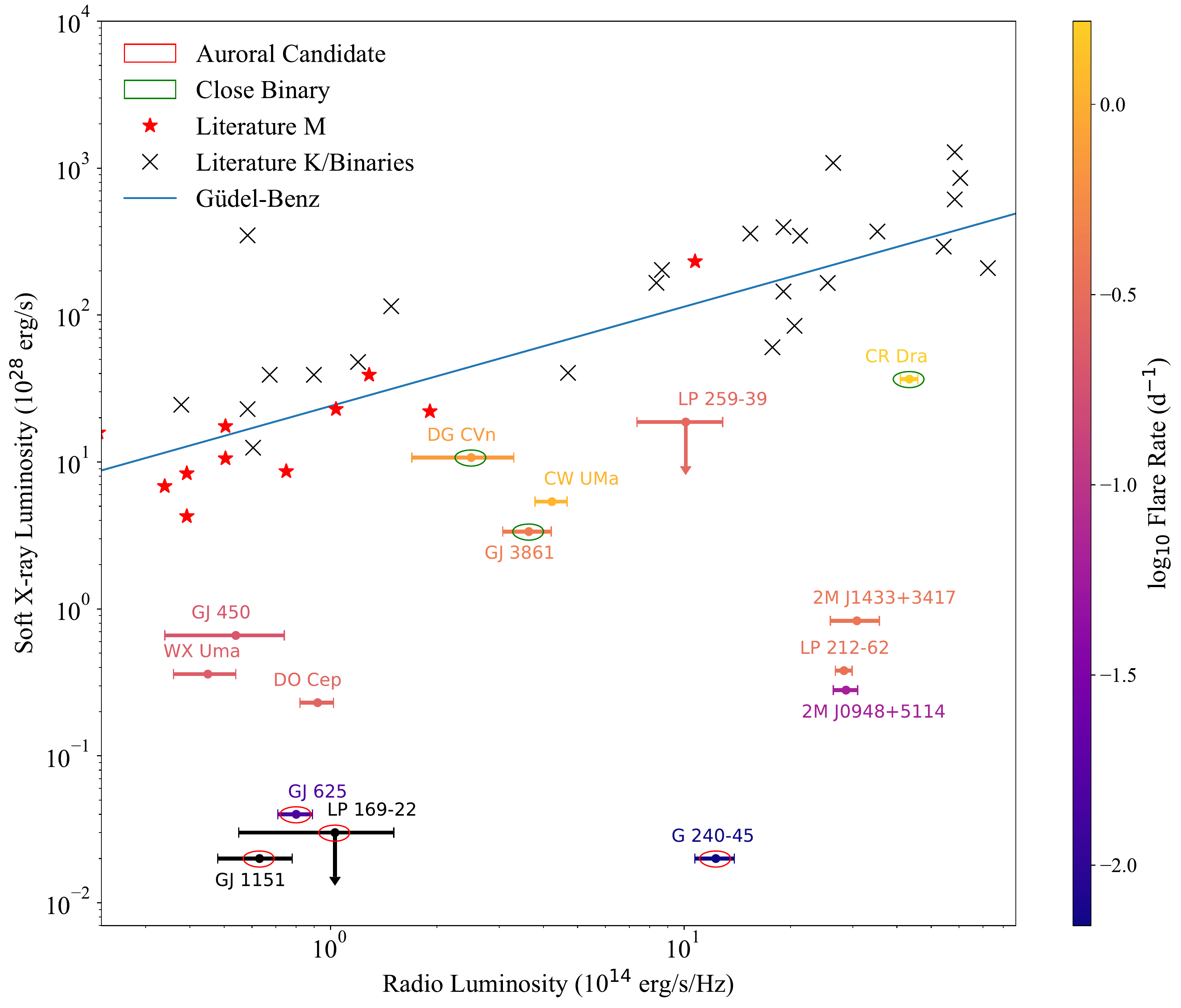}
\caption{Güdel-Benz diagram for all stars in our sample, colored by \textit{stella} flare rate, with individual stars highlighted.  Arrows denote upper limits. To show the context of the overall Güdel-Benz relation, we show literature sources as collated in \citet{callingham_pop}, using literature data from \citet{guedel93,benz94} together with a trend line. Red stars from denote M~dwarfs from literature, and black crosses denote K~dwarfs and interacting binaries. All stars in our sample are under-luminous in X-rays compared to the literature values, especially the auroral candidates which show two to three orders of magnitude lower flare rates and X-ray emission relative to similarly radio-bright stars. This figure was produced in a Jupyter Notebook, available online. \href{https://github.com/benjaminpope/mtv/blob/master/notebooks/paper_plots.ipynb}{\color{linkcolor}\faGithub} \label{gudel_benz}}
\end{figure}

We have therefore made an X-ray versus radio luminosity Güdel-Benz diagram for our sample (Figure~\ref{gudel_benz}), situating our stars relative to those from \citet{guedel93} and \citet{benz94} and coloring each LOFAR star by its TESS flare rate. All LOFAR stars fall below the Güdel-Benz relation, which as noted by \citet{callingham_pop} means that the majority are coherent emitters.

The four stars GJ\,625, GJ\,1151, G\,240-45, and LP\,169-22 are not only two to three orders of magnitude under-luminous in X-rays, but also two or more orders of magnitude lower in flare rate than comparably radio-bright sources. Their empirical flare rates would put all of them in the bottom 2.5\% of the M~dwarfs by flare rate determined by \citet{gunther20}. Their low flare rate and slow rotation favor an \ac{spi} model for radio emission.

The presence of flares does not in general indicate the absence of auroral emission. Young, rapidly-rotating M~dwarfs are more likely to flare \citep{feinstein20}, while the rapid ($\lesssim$2\,d) rotation period also allows radio emission via the breakdown of co-rotation. This means that these effects are confounded for active stars and cannot be disentangled based solely on these data. \citet{callingham_crdra} showed that CR\,Dra had an unusually high TESS flare rate among M~dwarfs, and detected both a quiescent component and a flaring component of radio emission. This quiescent component could be due to a breakdown of corotation in a magnetized plasma disk around one component of the binary system, and the flares may be due to local structure in the magnetosphere.

A cluster of actively-flaring stars with lower radio and X-ray emission includes WX\,UMa, DO\,Cep, and GJ\,450. Similar flare rates and X-ray luminosities are sustained by 2M\,1433+3417, LP\,212-62, and 2M\,0948+5114, but at much higher luminosities. 
While \citet{callingham_pop} proposed GJ\,450 may be quiescent due to its low H$\alpha$ emission, on the basis of its flare rate, we suggest that the radio emission from GJ\,450 and these other sources could plausibly be activity-driven or auroral.

DO\,Cep is one of the few stars from  \citet{callingham_pop} that was considered likely to be radio-bright due to plasma emission. In particular, DO\,Cep in unique among the LOFAR sample for its low circular polarization fraction ($38\pm5\%$). The location of WX\,UMa is harder to explain, as the radio emission is likely generated by ECMI \citep{davis2021}. The strong $\sim 3.5\,{\rm kG}$ surface dipole magnetic field of WX\,UMa \citep{morin10} plausibly allows ECMI to be produced via a large coronal loop, rather than breakdown of co-rotation or a satellite interaction. 


\section{Conclusions and Future Work}
\label{sec:conclusions}
 
TESS light curves reveal that the LOFAR radio emitting stars differ by orders of magnitude in optical flare rate. The least X-ray luminous sources are shown to have very low flare rates, which is further evidence that their emission is from \ac{spi} rather than activity-driven processes. 

Nevertheless, none of the \ac{spi} auroral emission candidates host known exoplanet candidates, or are TESS Objects of Interest transit candidates. Periodic radio emission, in phase with a known planetary period, would be definitive evidence of \ac{spi}.

We recommend further observations of GJ\,1151, LP\,169-22, G\,240-45, and GJ\,625 as high-priority LOFAR and \ac{rv} targets, to search for short-period planets that could cause this proposed \ac{spi}, or to sensitively rule out such planets. 

As shown by \citet{zic20}, simultaneous optical and radio flares can be used to determine the nature of both radio emission and optical signatures and probe the stellar space weather environment directly. Simultaneous LOFAR and TESS observations will be important in establishing the physical mechanism connecting optical and radio variability in these sources.
 
Short-period known planets around very quiet stars will also merit follow-up radio observations. For instance, \citet{nowak20} and \citet{cloutier20} have detected a pair of planets orbiting the M~dwarf LTT~3780, using both TESS photometry \ac{rv} instruments. The \citet{nowak20} team note especially that LTT~3780 is a similar star to GJ\,1151, and that given one planet has an ultra-short orbital period of 0.77\,d, it is a promising target for radio search. As noted by \citet{cloutier20} and reproduced by our pipeline, LTT~3780 is not found to flare in the entire TESS Sector~9 in which it is observed. We therefore encourage radio follow-up of this system and systems like it.

The low-frequency arm of the Square Kilometre Array, SKA-Low, will even in its Phase~1 have a sensitivity nearly an order of magnitude better than LOFAR in Stokes~V, and will detect many more stellar and \ac{spi} systems \citep{pope19radio}. The great majority of sources SKA-Low will detect will be in the Southern Hemisphere, and will be inaccessible to northern NIR instruments best suited to M~dwarf observations. We therefore recommend in the longer-term improving access to NIR \ac{rv} facilities in the Southern Hemisphere as part of SKA science and exoplanet science generally, as the best understanding of M~dwarf space weather and habitability will be gained from optical follow-up of SKA sources.

\section{Open Science}
\label{sec:open}

In the interests of open science, we have made available the Jupyter notebooks used to generate the figures in this paper, under a BSD 3-clause open source license at \href{https://github.com/benjaminpope/mtv}{github.com/benjaminpope/mtv}.  We encourage and welcome other scientists to replicate, apply, and extend our work.

\section*{Acknowledgements} 

BJSP and JRC are grateful to the School of Physics and Sydney Institute for Astronomy at the University of Sydney for hosting them as visiting researchers during the COVID-19 pandemic. We acknowledge and pay respect to the Gadigal people of the Eora Nation. It is upon their unceded, sovereign, ancestral lands that the University of Sydney is built. BJSP would like to acknowledge the traditional owners of the land on which the University of Queensland is situated, the Turrbal and Jagera people. We pay respects to their Ancestors and descendants, who continue cultural and spiritual connections to Country.

JRC thanks the Nederlandse Organisatie voor Wetenschappelijk Onderzoek (NWO) for support via the Talent Programme Veni grant. The LOFAR data in this manuscript were (partly) processed by the LOFAR Two-Metre Sky Survey (LoTSS) team. ADF acknowledges the support from the National Science Foundation Graduate Research Fellowship Program under Grant No.\,(DGE-1746045). MNG acknowledges support from MIT's Kavli Institute as a Juan Carlos Torres Fellow. Any opinions, findings, and conclusions or recommendations expressed in this material are those of the authors and do not necessarily reflect the views of the National Science Foundation. 

This team made use of the LOFAR direction independent calibration pipeline (\url{https://github.com/lofar-astron/prefactor}), which was deployed by the LOFAR e-infragroup on the Dutch National Grid infrastructure with support of the SURF Co-operative through grants e-infra 160022 and e-infra 160152 \citep{2017isgc.confE...2M}. The LoTSS direction dependent calibration and imaging pipeline (\url{http://github.com/mhardcastle/ddf-pipeline/}) was run on computing clusters at Leiden Observatory and the University of Hertfordshire, which are supported by a European Research Council Advanced Grant [NEWCLUSTERS-321271] and the UK Science and Technology Funding Council [ST/P000096/1]. This paper includes data collected by the TESS mission. Funding for the TESS mission is provided by the NASA Explorer Program. 

This research has made use of the SIMBAD database, operated at CDS, Strasbourg, France, and NASA's Astrophysics Data System. 

\software{This research made use of \textsc{stella} \citep{feinstein20joss, feinstein20}; \texttt{lightkurve}, a Python package for Kepler and TESS data analysis \citep{lightkurve}; the \textsc{IPython} package \citep{ipython}; \textsc{NumPy} \citep{numpy}; \textsc{matplotlib} \citep{matplotlib} \textsc{SciPy} \citep{scipy}; and Astropy, a community-developed core Python package for Astronomy \citep{astropy}; \textit{stella} \citep{feinstein20joss}}



\bibliography{ms}



\end{document}